\documentclass[aps,prl,reprint,tightenlines,superscriptaddress,amsmath,amssymb]{revtex4-1} 

\usepackage[utf8]{inputenc} % set input encoding (not needed with XeLaTeX)
\usepackage{graphicx} % support the \includegraphics command and options
\usepackage{natbib}
\usepackage{mathptmx}
\usepackage{xcolor}
\usepackage[colorlinks=true,hyperfigures=true,citecolor=blue,linkcolor=red]{hyperref}

\usepackage[english]{babel}
\usepackage{siunitx}
\newcommand{\thetaapp}{$\theta_{\rm\, app}$ }
\newcommand{\thetam}{$\theta_{\rm\, m}$ }

\DeclareSIUnit\pixel{px}

\begin{document}
\title{Spreading of granular suspensions on a solid surface}
\author{Menghua Zhao}
\author{Mathieu Ol\'{e}ron}
\author{Alice Pelosse}
\author{Laurent Limat}
\author{\'{E}lisabeth Guazzelli}
\author{Matthieu Roch\'{e}}
\email[E-mail address: ]{matthieu.roche@univ-paris-diderot.fr}
\affiliation{Mati\`{e}re et Syst\`{e}mes Complexes, CNRS UMR 7057, Universit\'{e} de Paris, 10 Rue A. Domon et L. Duquet, 75013 Paris, France}

\begin{abstract}
We examine the spreading of a suspension of non-Brownian spheres suspended in a Newtonian liquid on a solid substrate. We show that the spreading dynamics is well described by the classical Cox-Voinov law provided the value of the fluid viscosity that arises in the capillary number of the problem is adjusted to a value that depends on particle size and volume fraction in a non-trivial way. We demonstrate that this behavior is a signature of the ability of the particles to approach the contact line close enough to affect dissipation. 
\end{abstract}

\maketitle

\begin{figure}[ht]
\centering
\includegraphics[scale=1]{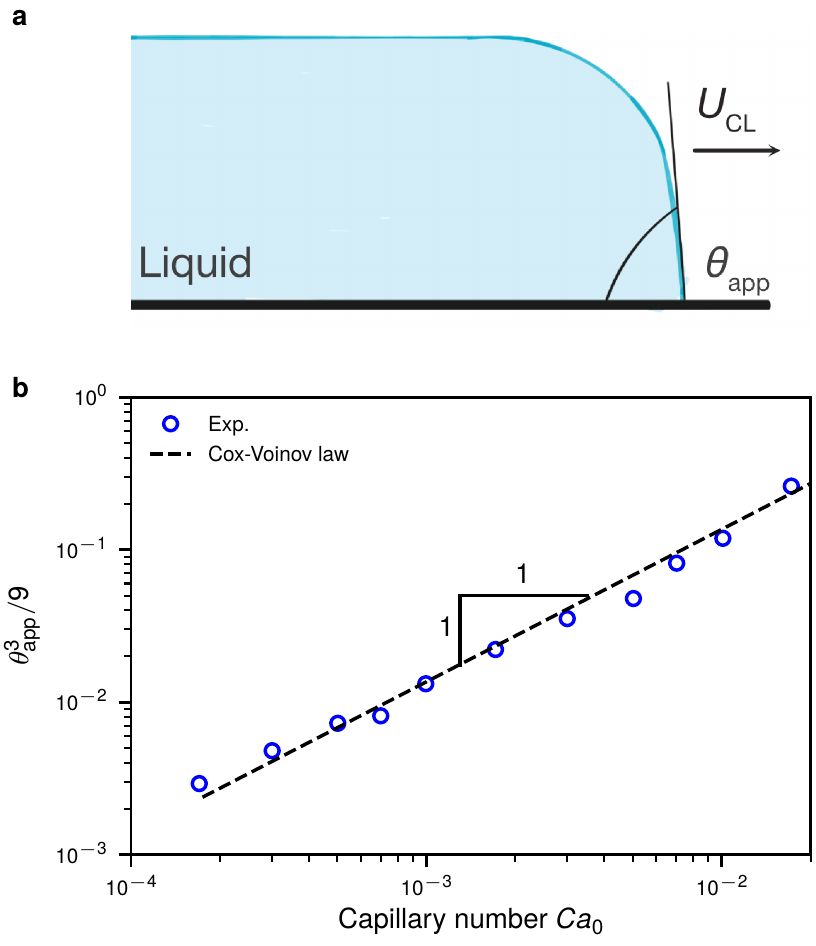}
\caption{(a) Sketch of the experimental set-up. (b) Comparison between experimental results for the apparent dynamic contact angle, $\theta_{\mathrm{app}}^3/9$,  as a function of the capillary number, $Ca_0$, for PEG-ran-PPG monobutyl ether and the Cox-Voinov solution, Eq.\,(\ref{eq:coxvoinov}).}
\label{fig:fig1}
\end{figure}

The control of wetting of liquids on surfaces is a topic of crucial importance to many industrial and natural systems over a large range of length scales, from protein folding and ion channel gating \cite{berne2009} to fibrous materials such as textiles and feathers \cite{rijke2011,duprat2012} \textit{via} ink-jet printing \cite{wijshoff2018}. As a consequence, wetting has attracted a lot of attention from  both fundamental \cite{Bonn2009,snoeijer2013} and engineering perspectives \cite{liu2017b}. One of the main challenges in this field is to understand how a liquid spreads on a surface, \textit{i.e.} how the contact line between the liquid, the solid, and the surrounding fluid relaxes to equilibrium (Fig.\,\ref{fig:fig1}a).

An essential feature of spreading flows is that their size decreases to zero at the contact line. This vanishing of length scale has fascinating physical consequences. First, macroscopic theories of spreading for simple liquids indicate that viscous stresses diverge at the contact line and prevent its motion, contrary to daily experience \cite{Huh1971}. This latter observation leads to the conclusion that the no-slip boundary condition at the interface between the liquid and the wall is violated at the contact line. Second, complex fluids have spreading properties that are hardly predictable based on their bulk properties. For example, the spreading of suspensions of nanoparticles, nanofluids, is enhanced because of particle layering at the contact line and the resulting disjoining pressure gradient \cite{wasan2003,wasan2011}. In these cases, confinement close to the contact line plays a significant role in setting the spreading dynamics.

Much less is known on the spreading of suspensions when the disjoining pressure is irrelevant and the  size of the particles $d_{\rm p}\gg$ \SI{1}{\micro\metre}, \textit{e.g.}\,for granular suspensions. Studies characterizing thin films deposited on plates extracted from baths of these fluids show that particles are entrained when the thickness of the film is commensurate with the particle diameter \cite{dimitrov1996,colosqui2013a,gans2019,palma2019}. Only very few studies have focused on the contact-line motion during the spreading of granular suspensions \cite{nicolas2005,mackaplow2006,han2012,kulkarni2016} despite its relevance to many cutting-edge technologies \cite{dimitrov1996,schneider2016,radcliffe2019}. Strong confinement effects are expected. In such a situation, the effective suspension viscosity is a non-trivial function of confinement \cite{fornari2016} and particle ordering may be observed \cite{dimitrov1996,colosqui2013a,gans2019,palma2019}. The effective-medium approach that describes well the suspension bulk flow \cite{guazzelli2018} breaks down and the discrete nature of the suspension must be accounted for.

The present contribution adresses the motion of the triple-phase contact line surrounding a droplet of granular suspension on a silicon wafer. The two main findings are that (i) the apparent contact angle depends on capillary number in the same way as for a pure liquid, but with an apparent viscosity which depend on both concentration and particle size unlike what is observed in bulk suspension rheology, and (ii) this discrepancy between the apparent viscosity extracted from wetting experiments and that obtained using bulk rheology results from the existence of a particle-depleted region at the contact line. We propose a model that extends the Cox-Voinov law to the case of granular suspensions using geometrical arguments and discuss its validity. In particular, it is found that the experimental observations cannot be rationalized on mere geometrical grounds.

%-------------------------------------------------------------------------------------------------------------------------------------------------
%:Materials and Methods
%-------------------------------------------------------------------------------------------------------------------------------------------------

The suspending fluid is polyethylene glycol-ran-propylene glycol monobutyl ether (PEG-ran-PPG ME, Sigma Aldrich, molecular weight $M_{\mathrm{w}}=3900$ \si{\gram\per\mole}, density $\rho_{\, 0}=$ \SI{1056}{\kg\per\cubic\meter}, surface tension $\gamma_{\mathrm{\, 0}}=$ \SI{35}{\milli\newton\per\meter}). Since the experiments were performed under ambient conditions (temperature  $20\leq T\leq30$ \si{\celsius}), the dependence of its shear viscosity $\eta_{\, 0}$ on temperature has been accounted for in the data analysis. The particles are spherical polystyrene beads (Microbeads Dynoseeds TS, average particle diameter $10 \leq d_{\rm p}\leq 550$ \si{\micro\meter}, density $\rho_{\rm\, p}=$ \SI{1050}{\kg \per\cubic\meter}). They are dispersed in PEG-ran-PPG ME in a beaker to obtain homogeneous neutrally-buoyant suspensions having volume fractions varying in the range $20\leq \phi \leq 40$\%. The surface tension of these suspensions has been found to be equal to that of the suspending liquid \cite{Couturier2011}. Their rheology is also well documented \cite{Boyer2011a,Couturier2011,Chateau2018}. In the range of volume fractions $\phi$ investigated, their bulk shear viscosity $\eta_{\rm\, s}$ increases with increasing $\phi$ while being independent of shear rate and particle size $d_{\rm p}$, before diverging at a maximum volume fraction $\phi_{\rm\, c}$. The dependence of $\eta_{\rm\, s}$ on $\phi$ is well represented by the Eilers empirical correlation \cite{guazzelli2018} :
\begin{equation}
\eta_{\mathrm{s}} (\phi)=\eta_0 \left[1+(5\phi/4)/(1-\phi/\phi_{\mathrm{c}})\right]^2.
\label{eq:eilers}
\end{equation}
\begin{figure}[ht]
    \centering
    \includegraphics[scale=1]{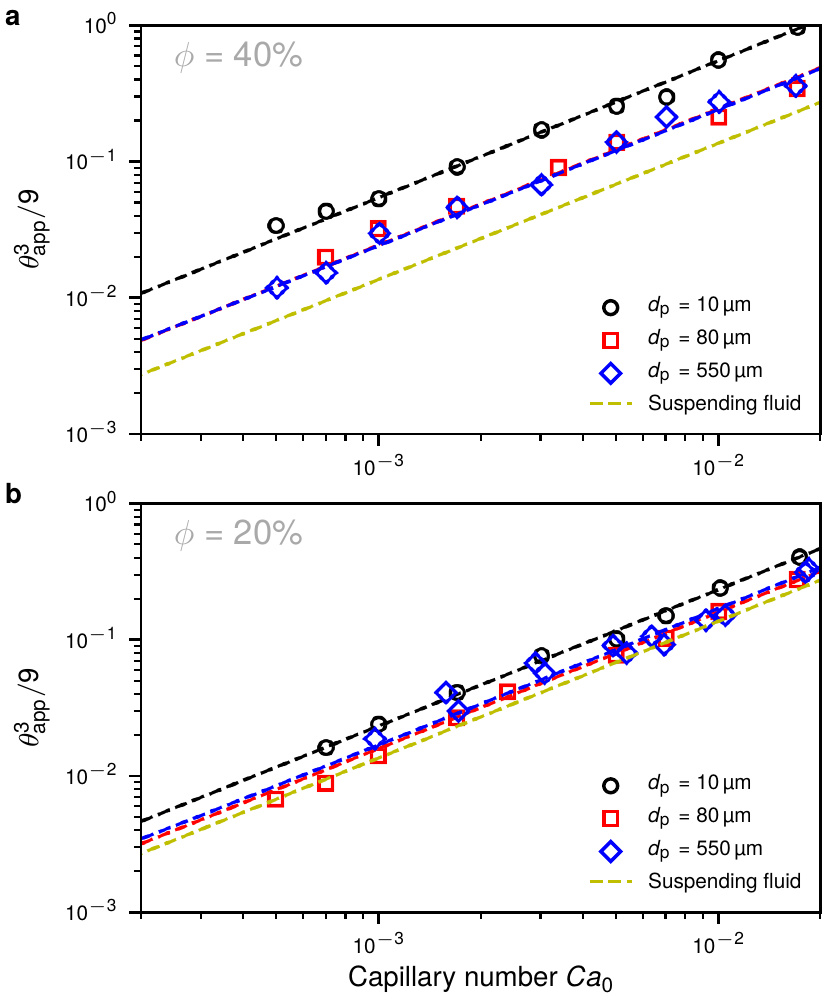}
    \caption{Dependence of $\theta_{\rm app}^3/9$ on the solvent capillary number $Ca_0$ for suspensions with particle volume fraction (a) $\phi=40$\% and (b) $\phi=20$\%. Yellow dashed line: Cox-Voinov solution for the pure solvent ($\phi=0$), Eq.\,(\ref{eq:coxvoinov}) with $Ca=Ca_0$. Analysis of suspension spreading is restricted to $4 \,10^{-4}\leq Ca_0\leq 2 \, 10^{-2}$ as reaching values out of this interval proved to be difficult.}
    \label{fig:fig2}
\end{figure}

A drop of suspension of typical volume \SI{300}{\micro\liter} is deposited  on the surface of a silicon wafer (diameter $d_{\rm w}=$ \SI{50}{\milli\metre}) using a steel needle (inner diameter $d_{\rm i}=$ \SI{4.58}{\milli\meter}) connected to a syringe mounted on a syringe pump (Fig.\,\ref{fig:fig1}a). The surface of the wafer is successively cleaned with acetone, ethanol, and DI water prior each experiment using a clean-room cloth. Side-view movies of suspension spreading are captured with a digital camera with a spatial resolution of \SI{30}{\micro\metre\per\pixel}. As the resolution of the camera is the same for all experiments, we measure the angles at the same location on the interface. Contact angles are measured with the software package FiJi \cite{Schindelin2012,Schneider2012} by adjusting manually a straight line to the air/liquid interface in the vicinity of the contact line. We run at least three experiments for each set of parameters and we find good reproducibility: the resulting uncertainty is smaller than the size of the marker in the graphs of Figs.\,\ref{fig:fig1} and \ref{fig:fig2}. We measure contact angles for axisymmetric drops. Top-view pictures or movies allow us to enforce the latter constraint and they are also used to characterize the suspension close to the contact line.  
\begin{figure*}[ht]
    \centering
    \includegraphics[scale=1]{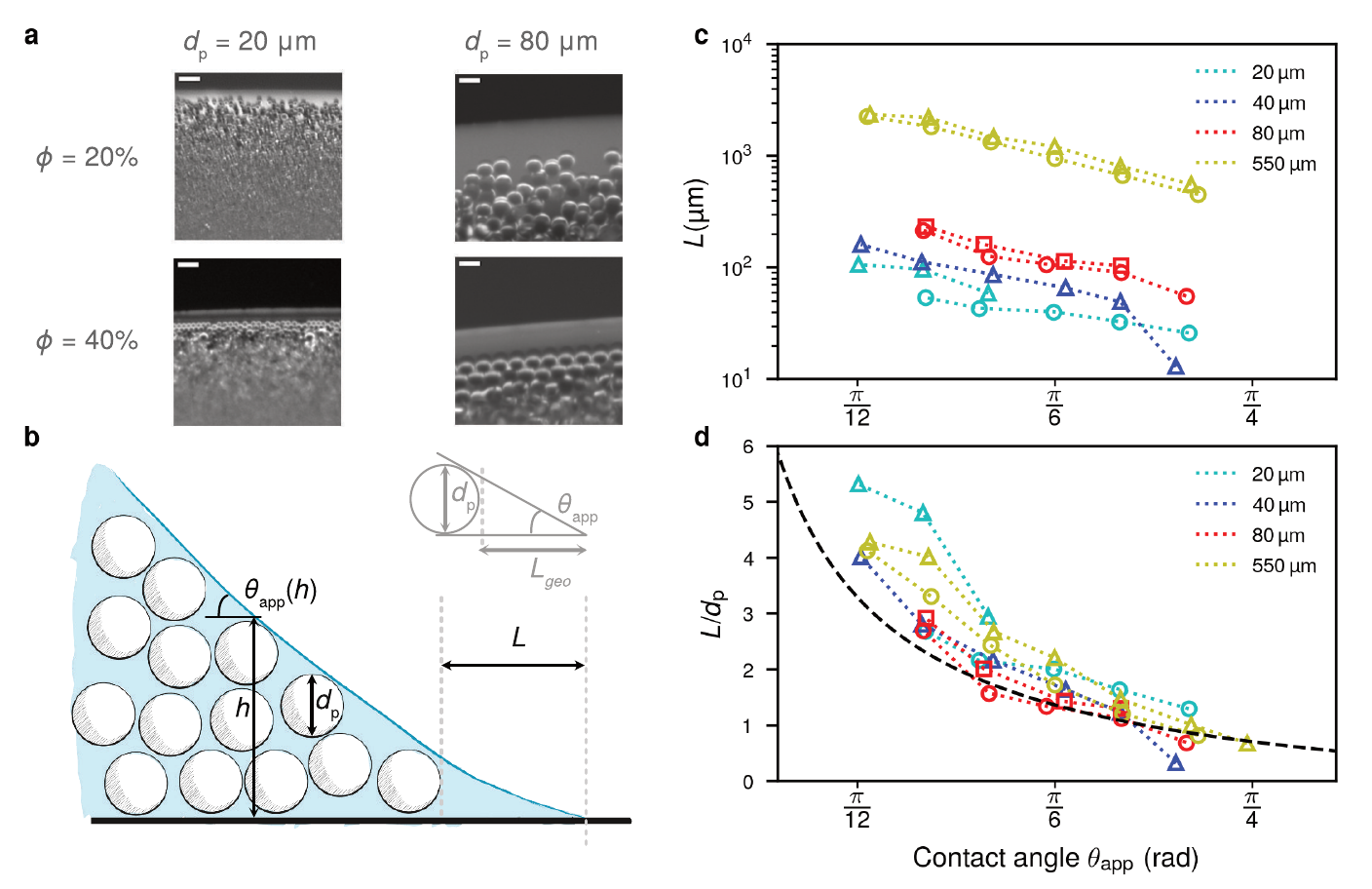}
    \caption{(a) Top-view images in the vicinity of the moving contact line for suspensions with two particle sizes, $d_{\rm p}=$ 20 and \SI{80}{\micro\meter}, and for two volume fraction, $\phi=20$\% and $\phi=40$\% (scale bar = \SI{100}{\micro\metre}). The brighter grey region between the contact line and the beads contains only the suspending liquid. (b) Sketch of the region around the contact line. The beads stay at a distance $L$ from the contact line. Inset: schematic of the geometry used to compute $L_{\rm\, geo}$, Eq. \ref{eq:geopredict}. (c,d) $L$ versus  $\theta_{\rm app}$ for $d_{\rm p}=20$, $40$, $80$, and \SI{550}{\micro\meter} in dimensional (resp. non-dimensional) form. $\phi=20$\% ($\square$), $\phi=30$\% ($\vartriangle$),$\phi=40$\% ($\circ$) In (d), black dashed line: minimal distance of a sphere of diameter $d_{\rm p}$ to the contact line, Eq.\,(\ref{eq:geopredict}).}
    \label{fig:fig3}
    \end{figure*}
%-------------------------------------------------------------------------------------------------------------------------------------------------
Studying the wetting of granular suspensions requires a prior characterization of the spreading of the pure suspending fluid (PEG-ran-PPG ME) on the silicon wafer. The prediction for the shape of the interface of the spreading drop in the vicinity of the contact line, known as the Cox-Voinov solution, can be written as:
\begin{equation}
\theta_{\mathrm{app}}^3=\theta_{\mathrm{m}}^3+9 Ca\ln{(h/\ell)},
\label{eq:coxvoinov}
\end{equation}
with \thetaapp the apparent dynamic contact angle, \thetam the microscopic contact angle at $h=\ell$, $Ca=\eta U_{\mathrm{CL}}/\gamma$ the capillary number of the system based on the kinematic viscosity $\eta$ and the surface tension $\gamma$ of the liquid as well as the contact line velocity $U_{\rm CL}$, $\ell$ a nanoscopic cut-off scale that acts as a slip length and that is meant to circumvent issues with stress singularity, and $h$ the height inside the droplet where \thetaapp is measured (Note that, following the original derivation of Voinov \cite{voinov1976}, Eq.\,(\ref{eq:coxvoinov}) provides the $h$-dependence of \thetaapp but the more common $x$-dependence \cite{Bonn2009} can be recovered by assuming $h\approx$ \thetam $x$, $x=0$ being the location of the contact line.). We define $Ca_{\mathrm{0}}$ as the capillary number obtained using the properties of PEG-ran-PPG ME and the experimental velocity of the contact line $U_{\rm\, CL}$. As the apparent equilibrium contact angle of PEG-ran-PPG ME on silicon is $\theta_{\rm eq}\sim$ \SI{6}{\degree} and we assume that $\theta_{\rm\, m}=\theta_{\rm\, eq}$, we are almost in a situation of total wetting. Thus we expect the $\theta_{\mathrm{app}}^3(Ca_{\mathrm{0}})$ curve for the suspending liquid to be a line of slope 1 in log-log representation. Fig.\,\ref{fig:fig1}b shows that this expectation is fulfilled and that Eq.\,(\ref{eq:coxvoinov}) reproduces well the present observations with a value for $\ln(h/\ell)\approx \ln(\theta_{\rm eq} x/\ell) \sim 13.76$. 

We now turn to the spreading of granular suspensions. Assuming Eq.\,(\ref{eq:eilers}) valid and $\ell$ identical to the value found for the suspending liquid, we anticipate a shift of the $\theta_{\mathrm{app}}^3(Ca_{\mathrm{0}})$ curves towards higher values of \thetaapp for suspensions when compared to the case of the pure suspending liquid. Eq.\,(\ref{eq:eilers}) also suggests that the shift should be independent of particle size, $d_{\rm p}$. Fig.\,\ref{fig:fig2} shows that the situation is more complex than expected. For all the suspensions that were tested, the data align along lines parallel to that obtained for the solvent. There is a conspicuous shift for suspensions with the largest particle volume fraction $\phi=40$\% (Fig.\,\ref{fig:fig2}a). However, the shift decreases as the size of the particles increases, in contrast with our initial expectation. The suspension data even collapse on the curve for the pure suspending fluid for the largest particles at a volume fraction $\phi=20$\% (Fig.\,\ref{fig:fig2}b).

Top-view imaging of the moving contact line sheds light on the reason behind the particle-size dependence of the apparent shear viscosity of suspensions during spreading. Figure \ref{fig:fig3}a shows that the beads remain at a finite distance $L$ from the contact line. There is a pure-fluid region devoid of particles in the vicinity of the contact line (Fig.\,\ref{fig:fig3}b). A striking feature is that the first layers of beads outside the depleted region are crystallized for drops with the largest volume fraction $\phi=40$\%, much like what is observed for nanofluids \cite{wasan2003}. A similar ordering is also observed at lower volume fractions with an extent decreasing with decreasing $\phi$. As the height of the air/liquid interface increases further, the beads switch from a crystal-like to a disordered structure. These observations show that the particle concentration field is singular at $x=L$, $\nabla\phi(x=L)\rightarrow\infty$.

The extent of the depleted region $L$ decreases with increasing  \thetaapp but increases with increasing $d_{\rm p}$ (Fig.\,\ref{fig:fig3}c). The dependence on $\phi$ is weak. Rescaling $L$ by $d_{\rm p}$ leads to a good, albeit imperfect, collapse of the datasets (Fig.\,\ref{fig:fig3}d). The demixing in the vicinity of the contact line seems thus to be linked to the inability of particles to flow inside the contact-line corner for height typically smaller than their particle diameter. A geometric description of the depletion in particles is then tempting. For the sake of simplicity, we assume that the air/suspension interface is rigid and we neglect its curvature (Inset in Fig. \ref{fig:fig3}b). The minimal geometric distance, $L_{\rm geo}$, that a sphere can reach in a corner having an angle \thetaapp is:
\begin{equation}
    L_{\rm\, geo}/d_{\rm p}=\frac{1}{2}\left [\left(\tan{(\theta_{\rm app})}^{-1}+\sin{(\theta_{\rm app})}^{-1}\right)-1 \right ].
    \label{eq:geopredict}
\end{equation}
This scaling captures the intuitive results that $L_{\rm\, geo} \rightarrow\infty$ as $\theta_{\rm app}\rightarrow 0$ and $L_{\rm\, geo} \rightarrow 0$ as $\theta_{\rm app}\rightarrow \pi/2$.  The prediction of Eq.\,(\ref{eq:geopredict}) provides a lower bound to the experimental datasets (Fig.\,\ref{fig:fig3}d). The observations reported in Fig.\,\ref{fig:fig3} show the crucial importance of the discrete nature of the suspension and of the confinement of particles in the vicinity of the contact line. 

As a first attempt at capturing these observations, we propose a model based on the confinement-induced depletion of the particles close to the contact line. We assume that the flow in this region results from the matching of two corner flows (Fig.\,\ref{fig:fig3}b), one in the depleted region (\textit{i.e.}\,$h\leqslant d_{\rm p}$) and the other in the suspension region (\textit{i.e.}\,$h \geqslant  d_{\rm p}$). Two microscopic cut-off lengths appear in this treatment,  $\ell$ in the pure-liquid corner and $d_{\rm p}$ in the particle-dense region. The slope of the air/liquid interface in both regions can then be written as:
\begin{align}
\theta_{\rm app}^3&=\theta_{\rm eq}^3+9Ca_0\ln{(h/\ell)} &\mbox{for}\; h\leqslant d_{\rm p},
\label{eq:coxvoinov0}\\
\theta_{\rm app}^3&=\theta_{d_{\rm p}}^3+9Ca_s\ln{(h/d_{\rm p})} & \mbox{for}\; h\geqslant d_{\rm p},
\label{eq:coxvoinovS}
\end{align}
with $Ca_{\rm s}=\eta_{\rm\, s}(\phi) U_{\rm\, CL}/\gamma= Ca_{0} \eta_{\rm\, s}(\phi)/\eta_{\, 0}$ and $\theta_{d_{\rm p}}$ in Eq.\,\ref{eq:coxvoinovS} the angle at $h=d_{\rm p}$. Imposing continuity of the slope of the air/liquid interface at $h=d_{\rm p}$ leads to:
\begin{equation}
\theta_{\rm\, app}^3=\theta_{\rm\, eq}^3+9
\left[ 
\ln{(\frac{h}{\ell})}
+\frac{\eta_{\rm\, s}(\phi) - \eta_{\, 0}}{\eta_{\, 0}}\ln{(\frac{h}{d_{\rm p}})}
\right]Ca_0.
\label{eq:coxvoinovAS}
\end{equation}
The quantity within the square brackets contains two terms that account for dissipation between scales $h$, $\ell$ and $d_{\rm p}$. 

%An important outcome of the model is that the value of $\theta_{\rm app}$ measured for a suspension at a given value of $Ca_0$ should be identical to that of the suspending liquid when $h\leqslant d_{\rm p}$.

The model provides a prediction for the amplitude of the size-dependent translation of the curves $\theta_{\rm\, app}^3/9$ versus $Ca_0$ observed for suspensions in Fig.\,\ref{fig:fig2}. While the slope is $\ln{(h/\ell)}$ for the pure PEG-ran-PPG ME, it is $(\eta_{\rm s,app}/\eta_0)\ln{(h/\ell)}$ for suspensions with an apparent viscosity of the suspension, $\eta_{\rm s,app}$, inferred from Eq.\,(\ref{eq:coxvoinovAS}) to be:
\begin{equation}
    \frac{\eta_{\rm\, s,app}}{\eta_0}=1+\frac{\eta_{\rm\, s}(\phi)-\eta_0}{\eta_0}\frac{\ln{(h/d_{\rm p})}}{\ln{(h/\ell)}}.
    \label{eq:apparentspreadingviscosity}
\end{equation}
Eq.\,(\ref{eq:apparentspreadingviscosity}) indicates that $\eta_{\rm\, s,app}\rightarrow \eta_0$ if $\theta_{\rm app}$ is measured at a height $h \rightarrow d_{\rm p}$. However, when $h \leqslant d_{\rm p}$, the h-dependent term on the r.h.s. becomes unphysical as the particle term decreases the apparent viscosity down to negative values that can become infinite.

\begin{figure}[t]
    \centering
    \includegraphics[scale=1]{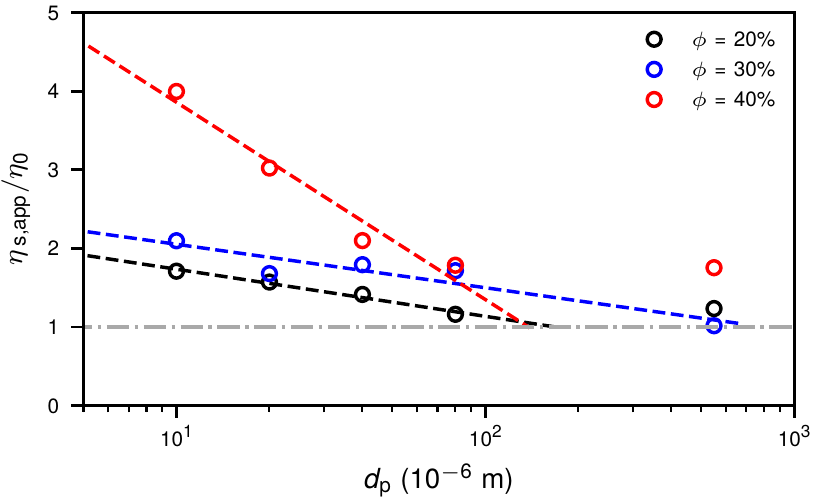}
    \caption{Dependence of the relative effective viscosity of granular suspensions during spreading $\eta_{\rm s,app}/\eta_0$ on particle diameter, $d_{\mathrm{p}}$. Dashed lines: fits of the experimental data by Eq.\,(\ref{eq:apparentspreadingviscosity}). The fitting parameters are (h=\SI{168}{\micro\metre}, $\eta_{\rm\, s}(\phi)=4.3\eta_{\, 0}$) for $\phi=$ 20\%, (\SI{799}{\micro\metre}, $4.6\eta_{\, 0}$) for $\phi=$ 30\% and (\SI{138}{\micro\metre}, $16\eta_{\, 0}$) for $\phi=$ 40\%. Dot-dashed line: physical bottom limit to the relative viscosity.}
    \label{fig:fig4}
    \end{figure}
 
We attempted at testing the model against experiments using values of viscosity given by Eq.\,\ref{eq:eilers}.  However, as the fits were completely unsatisfactory, we resolved to leave the suspension viscosity as a free parameter. The values of $\eta_{\rm\, s,app}/\eta_0$ extracted from the data were then fitted by Eq.\,(\ref{eq:apparentspreadingviscosity}) in Fig.\,\ref{fig:fig4}, leaving the height $h$ at which the angle is measured and the bulk apparent shear viscosity $\eta_{\rm\, s}(\phi)$ as adjustable parameters. The obtained fits happen to capture well the trends of the datasets and indicate that $h=\mathcal{O}(150)$ \si{\micro\meter}, except at $\phi=30$\% for which $h=$ \SI{799}{\micro\meter}. This order of magnitude for $h$ is sensible since we measure the angle over a few pixels. However, other severe disagreements are found. First, $\eta_{\rm\, s,app}/\eta_0$ differs from 1 for $h/d_{\rm p}\ll1$ (in particular for large $\phi$). Second, and most importantly, the values obtained for $\eta_{\rm\, s}$, and as a consequence for $\phi_{\rm\, c}$, are unrealistic. While the maximum flowable volume fraction is $\phi_{\rm\, c}\approx 54-58$\% for the present suspensions \cite{Chateau2018}, the fits point to different values of $\phi_{\rm\, c}$ for each $\phi$, from 47\% for $\phi=40$\% down to 25\% for $20$\%. 

In conclusion, the investigation of the spreading of granular suspensions on a solid surface indicates that the apparent viscosity of these suspensions in the vicinity of the triple-phase contact line is dependent on particle size. This observation results from the existence of a particle-depleted region at the contact line from which beads have been expelled due to confinement. A simple model attempting at a geometrical description of the particle depletion is however not sufficient to encompass the spreading dynamics. Further modeling would need to consider the transition region between the particle-depleted region and the dense region where ordering can occurs and how the viscosity of the suspension varies as a function of the height of the flow. The complex dependence on the confinement parameter that has been seen \textit{e.g.}\,between two plates \cite{fornari2016} may also play some role even though here one of the confining walls in a free deformable surface. Accounting for the strong particle displacement correlations and particle contact forces when suspensions becomes dense \cite{guazzelli2018} may also be important. Finally, we would like to point out that the logartihmic term $\ln{(h/d_{\rm p})}$ plays a crucial role here, as these two lengthscales are of comparable magnitude. This situation differs from that of simple fluids, as $h\gg\ell$ and the log term has almost constant magnitude given the accessible spatial resolution. Therefore the height at which the contact angle is measured in the case of granular suspensions must be carefully reported.
\begin{acknowledgments}
ANR (Agence Nationale de la Recherche) and CGI (Commissariat \`{a} l’Investissement d’Avenir) are gratefully acknowledged for their financial support of this work through Labex SEAM (Science and Engineering for Advanced Materials and devices), ANR-10-LABX-0096 and ANR-18-IDEX-0001.
\end{acknowledgments}
\bibliographystyle{apsrev}
\bibliography{suspread}

\end{document}